# Controlling Gaussian and mean curvatures at microscale by sublimation and condensation of smectic liquid crystals

*Dae Seok Kim[1], Yun Jeong Cha[1], Mun Ho Kim[2], Oleg D. Lavrentovich[3,*], Dong Ki Yoon[1,*]*

*To whom correspondence should be addressed.

E-mail: nandk@kaist.ac.kr and olavrent@kent.edu

[1] Graduate School of Nanoscience and Technology and KINC, Korea Advanced Institute of Science and Technology, Daejeon 305-701, Korea

[2] Department of Polymer Engineering, Pukyong National University, Busan, 608-739, Republic of Korea

[3] Liquid Crystal Institute and Chemical Physics Interdisciplinary Program, Kent State University, Kent, Ohio 44242, USA

# Abstract

Soft materials with layered structure such as membranes, block copolymers, and smectics exhibit intriguing morphologies with nontrivial curvatures. We report on restructuring the Gaussian and mean curvatures of smectic A films with free surface in the process of sintering, i.e. reshaping at elevated temperatures. The pattern of alternating patches of negative, zero, and positive mean curvature of the air-smectic interface has a profound effect on the rate of sublimation. As a result of sublimation, condensation, and restructuring, initially equilibrium smectic films with negative and zero Gaussian curvature are transformed into structures with pronounced positive Gaussian curvature of layers packing, seldom seen in samples obtained by cooling from the isotropic melt. The observed relationship between the curvatures, bulk elastic behaviour, and interfacial geometries in sintering of smectic liquid crystals paves the way for new approaches to control soft morphologies at micron and submicron scales.

**M**orphogenesis, defined as the ensemble of mechanisms responsible for the formation of patterns and shapes, is a fundamental issue in natural sciences in general[1] and physics of soft matter in particular. Especially profound examples of intimate relationship between the molecular structure of a soft material and its bulk curvatures and shapes are found in layered systems with a one-dimensionally periodic arrangement of molecules, such as block copolymers, stacked membranes, and various liquid crystals. These materials show high flexibility and shape adaptation, controlled by equidistance of the layers and by balance of the interfacial and layers bending energies.[2]

In the simplest realization, the so-called smectic A (SmA), the layer thickness is defined by the length of rod-like molecules. Smectics with soft boundaries, bordering either an isotropic fluid[3-11] or air[12-18], exhibit a very broad range of interesting morphologies. The well-known examples are highly irregular SmA nuclei emerging from the isotropic melt with layers bent into the arrays of focal conic domains[3] and smectic films with free surface that exhibit non-flat interfaces with depressions[12,19,20] and grooves[21,22]. Recently, new types of hierarchical assemblies have been added for SmA films with modulated substrates serving as topological templates.[15-18,23]

An intriguing question addressed in our work is how the shape of a SmA with a soft interface would change when the constituent molecules can leave the sample and re-condense on it. Such a system is a semifluorinated SmA with a free surface: the layers can be peeled off through sublimation at elevated temperatures.[24] From the studies of powdered metals and ceramics, it is known that sintering, defined broadly as a change in shape of a material held at elevated temperatures, is driven by the mean curvature of the interface that determines the local vapour pressure [25,26]. SmA features anisotropic surface tension and flexible layers that can bend.

We demonstrate that thermal treatment of SmA films leads to profound changes of their morphology, altering shape, mean and Gaussian curvature of the layers. The emerging structures are dramatically different from sintered crystalline and amorphous materials and from equilibrium SmA samples. The richness of the morphological changes is associated with a variable curvature of the SmA-air interface; this variable curvature mediates the local sublimation and condensation rates.

## Results

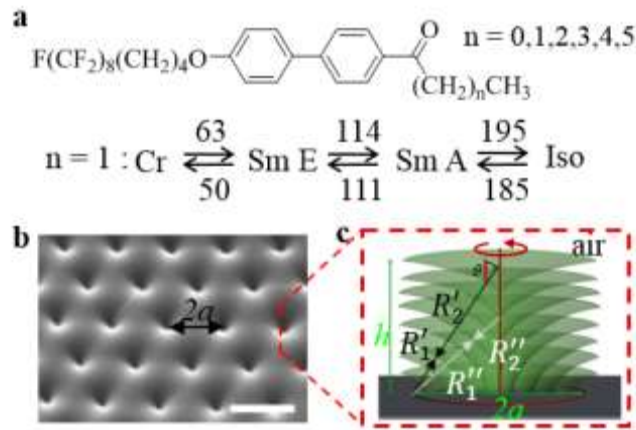

**Figure 1. Materials used and SmA film with TFCDs array** (a) Chemical structure and phase diagram of semifluorinated **Y002**. (b) Equilibrium structure of a SmA film surface viewed under the SEM with a hexagonal array of indentations at the apexes of TFCDs. The scale bar is 10 μm. (c) TFCD structure; $R_1'$, $R_2'$, $R_1''$, and $R_2''$ are the principal radii of layers curvature. The SmA molecules are everywhere oriented along the straight lines such as $R_1' R_2'$ and $R_1'' R_2''$ that join the points of axial defect with the points of circular base. The structure is axially symmetrical, as indicated by the curved arrowed loop. The Gaussian curvature is everywhere negative; the mean curvature can be either positive (as for the point with the principal radii of curvature $R_1'$ and $R_2'$) or negative (the point with $R_1''$ and $R_2''$). The angle $\theta$ is between the normal to the bottom substrate and the normal to the smectic layer (which is also the director).

**Preparation and characterization of SmA film with TFCDs** The SmA film was prepared by placing a droplet of material Y002 (n=1), Fig. 1a, in its isotropic melted state onto a silicon substrate treated with polyethyleneimine (PEI) and then cooling it down below 195 ℃ into the smectic state, Supplementary Table 1. Y002 spreads into a flat film of thickness $h = 10$ - $20\ \mu m$. The SmA molecules prefer to align parallel to PEI surface and perpendicularly to the free surface. SmA is in contact with air in all the experiments. Because of the antagonistic boundary conditions, the layers are curved in the vertical plane, with a radius of curvature $R_1$ in Fig. 1c. To avoid wall defects, the layers are additionally curved in the horizontal plane, with a radius of curvature $R_2$ originating at the axis of rotational symmetry in Fig. 1c. The folded layers form the so-called toroidal focal conic domain (TFCD). Figure 1c shows, as an example, two pairs of principal radii of curvature: the pair $(R_1', R_2')$ characterizes a point at a smectic layer that is close to the periphery of the TFCD, while $(R_1'', R_2'')$ corresponds to a point close to the vertical axis of rotational symmetry of the TFCD. In each pair, the directions of the radii are antiparallel to each other; the radii originate at two linear defects, namely, at the horizontal circle of radius $a$ (radii $R_1'$ and $R_1''$), and at the vertical axial line ($R_2'$ and $R_2''$). At the core of these two defects, the translational order of smectic layering is broken.

To summarize, within each TFCD, the layers are of a saddle-like shape with a negative Gaussian curvature, $G = 1/(R_1 R_2) < 0$.[12-14] Each TFCD is smoothly embedded into the surrounding system of flat horizontal layers, $G = 0$, as the layers cross the cylindrical lateral boundary of the TFCD in a perpendicular fashion, Fig. 1c. The smectic-air interface has a non-flat profile with cusp-like depressions at the axial defect in the centre of each TFCD, Fig.1b. When the smectic film with a hexagonal array of TFCDs is sintered at elevated temperatures within the range of stability of the SmA phase, it experiences dramatic restructuring, the result of which depends on the temperature and duration of sintering. Figure 2 shows examples of the most frequently met sintered structures.

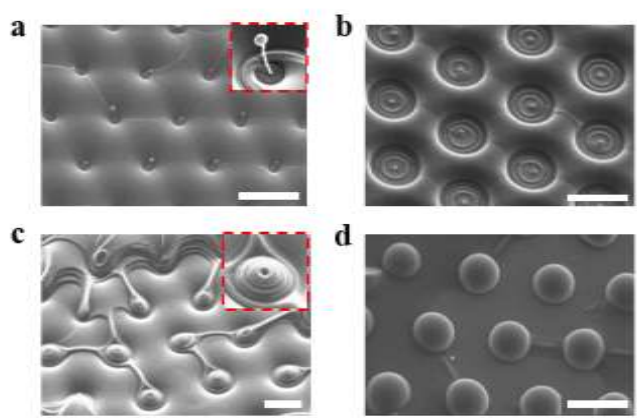

**Figure 2. SEM images of SmA films transformed by sintering (a)** Udumbara flower-like morphologies produced by sintering at 130 ℃ for 120 h. **(b)** Concentric ring structures produced by sintering at 160 ℃ for 40 min and **(c)** at 180 ℃ for 8 min. **(d)** Domes produced by sintering at 190 ℃ for less than 2 min. All scales are 10 µm.

**Sintering at 130 ºC** A sample was kept at 130 ºC for 120 h in air. Sublimation of the original film with a hexagonal array of indentations, Fig. 1b, results in partial removal of the smectic layers, Fig. 2a and Supplementary Figs. 1a,b. The edges of remaining partial layers are seen in Fig. 2a as sharp bright curved lines connecting the interiors of the neighboring TFCDs. An intriguing feature of the resulting morphology is an Udumbara flower-like formation, representing a vertical stalk centred at the symmetry axis of the original TFCD and capped by a tiny sphere (Fig. 2a). Other fluorinated SmA materials also show Udumbara flower-like structures, Supplementary Figs. 1c,d. The stack itself represents nested pieces of SmA layers, Supplementary Fig. 1c, or a linear array of toroidal and bead-like formations, Supplementary Fig. 1d. The region around the stalk, closer to the TFCD periphery, shows a much stronger degree of sublimation; removal of the material in this region is deeper than in the interstitial regions between the TFCDs with flat SmA layers.[24]

**Sintering at 160 ºC** The sample was kept at 160 ºC for 40 minutes. Sublimation removes most of the TFCDs and reveals the circular base of the domains, Fig. 2b, decorated by concentric rings. The rings are formed by many curved SmA layers, being much wider (100 nm - 1 $\mu$m) than the single layer. The vertical cross-section of the ring (Supplementary Fig. 2) is close to a hemi-circle, the centre of which is slightly displaced below the substrate surface. The downward shift might be caused by the change of anchoring[27] or by condensation of the material at the edges of the hemi-tori that would produce incomplete layers, Supplementary Fig. 2. If the sintering time is shorter than 40 min, then the removal of curved layers within the TFCD is only partial, similarly to the case described below.

**Sintering at 180 ºC** Sintering for about 6-10 min results in partial peeling of the layers. The layers are removed predominantly from the periphery of each TFCD and, to a somewhat lesser degree, from the interstitial regions between TFCDs, Fig. 2c, with a progressive loss of birefringence in the textures, Supplementary Fig. 3. As a result, the profile of the SmA film is formed by a hexagonal array of conical pyramids centred at the axes of the original TFCDs, Supplementary Fig. 4. Similar features of an accelerated sublimation at the periphery of TFCD and preserved central conical region are also evident in the textures reported previously.[24] The side surface of the conical pyramids exhibits hemi-tori, similar to rings observed in Fig. 2b and Supplementary Fig. 2, in which the concentric SmA layers curve to maintain equidistance and perpendicular orientation of molecules at the SmA-air interface.

**Sintering at 190 ºC** Sintering near the melting point for about less than 2 minutes results in a very fast restructuring that generates dome-like structures (Fig. 2d). The film loses roughly 90% of its volume, since the average volume of each TFCD is on the order of $10^3$ $\mu m^3$, while the volume of each dome is about $10^2$ $\mu m^3$, Supplementary Fig. 5. The domes have a circular base, resembling a portion of sphere (Fig. 2d, Supplementary Fig. 6c and Supplementary Fig. 7),

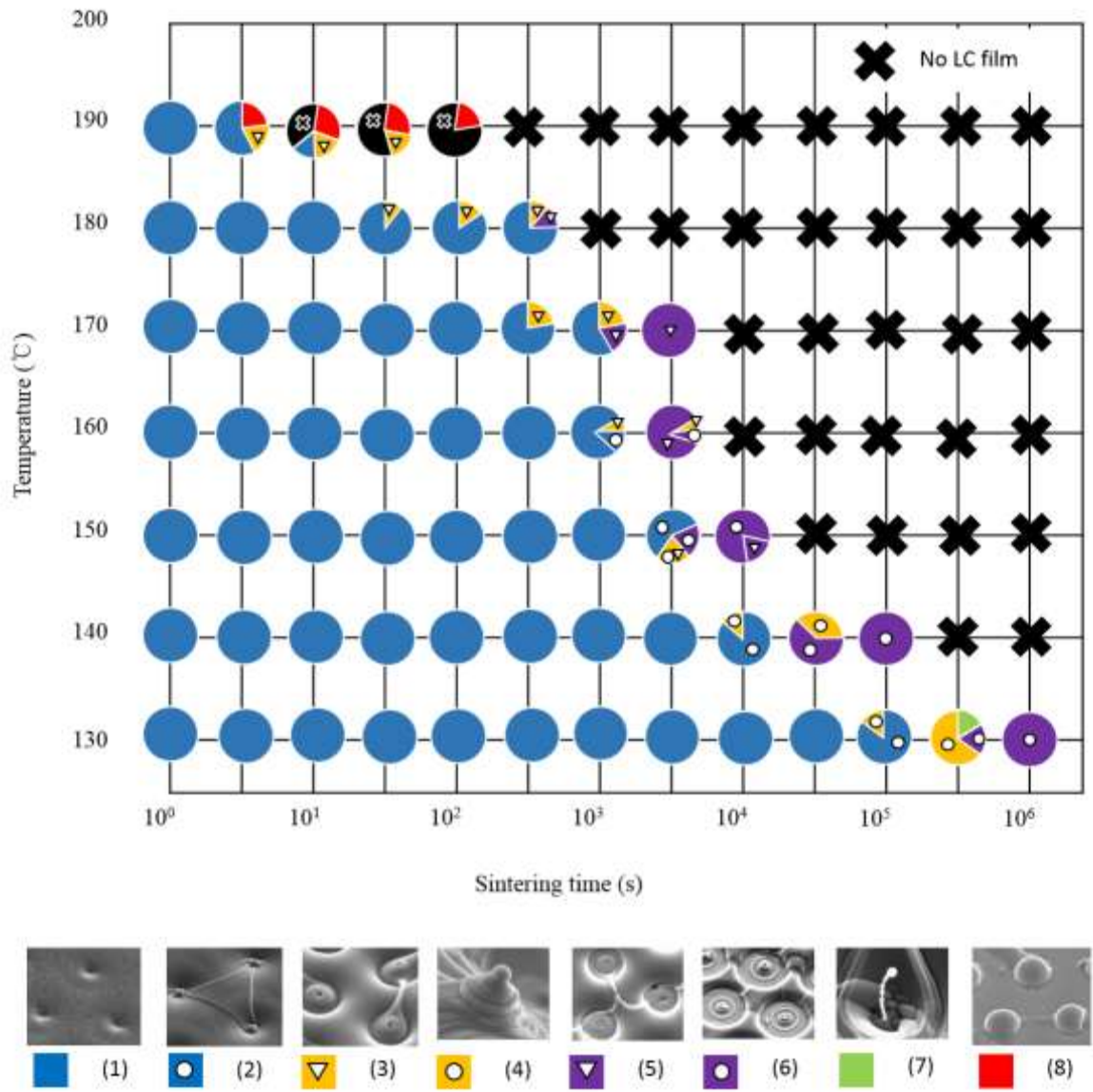

**Figure 3. Sintering diagram of morphological transformations of smectic films with TFCDs**
Eight distinct types of resulting structures are identified as a function of the sintering temperature and duration : (1) TFCDs with cusp-like depressions at the axes; (2) TFCDs with spherical caps at the axes; (3) conical pyramids with cusp-like depressions at the apex; (4) conical pyramids with spherical caps; (5) concentric rings with central cusp-like depression; (6) concentric rings with spherical caps at the centre; (7) Udumbara flowers-like structures; (8) domes. In the diagram, the eight types are marked by coloured segments with the area approximately corresponding to the frequency of the observed structures for a given set of (temperature, time) conditions. The eight types are illustrated by the corresponding SEM textures.

although sometimes one also observes fat torus-like domes, with an indentation in the centre,

Supplementary Fig. 6b. Atomic force microscope (AFM), Supplementary Fig. 7, and SEM,

Supplementary Fig. 5, shows that the domes are not exactly hemi-spherical. As in the case of rings, Fig. 2b, the reason might be a modified anchoring[27] or capillary condensation at the edges of domes, Supplementary Fig. 5b.

**Sintering diagram of morphological transformation of SmA films** In order to demonstrate the full set of scenarios and reproducibility of the results, the experiments were performed in the entire temperature range between 130 °C and 190 °C with a 10 °C increment, for different sintering times, ranging from 5 to $10^6$ seconds. For each temperature-duration condition, the experiments were performed at least 3-10 times, with qualitatively the same outcomes. The resulting diagram, Fig.3 and Fig.4, reveals eight distinct types of structures created by sintering:

(1) TFCDs with cusp-like depressions at the centre, Fig.4m; these are not much different from the original hexagonal arrays of TFCDs, Fig.1b, and are observed when the sintering time is short and the sintering temperature is low;

(2) TFCDs with spherical caps instead of the depressions formed at relatively low temperatures (130-160) °C and relatively short sintering times, Fig.4n,m,p,s,t. Figure 5 illustrates the process of the cap formation. First, the uppermost layer of the material sublimates, ruptures and shrinks; its edge changes from a curvilinear triangle, Fig.5a, to an oval, Fig.5b, and then a circle, Fig.5c. At the final stage, the shrinking layer closes up by acquiring a spherical cap, Fig.5d, presumably via re-condensation of SmA molecules from the air and diffusion within the SmA bulk. Diffusion should be facilitated at the core of the central defect where the layered structure of SmA is broken.

(3) Conical pyramids with cusp-like depressions at the centres (Fig.2c, Fig.4d,e,f,g,h,i,j,k,l) formed at high temperatures (150-190) °C and low-to-moderate sintering times;

(4) Conical pyramids with spherical caps at the centres (Fig.4q,t), formed at temperatures

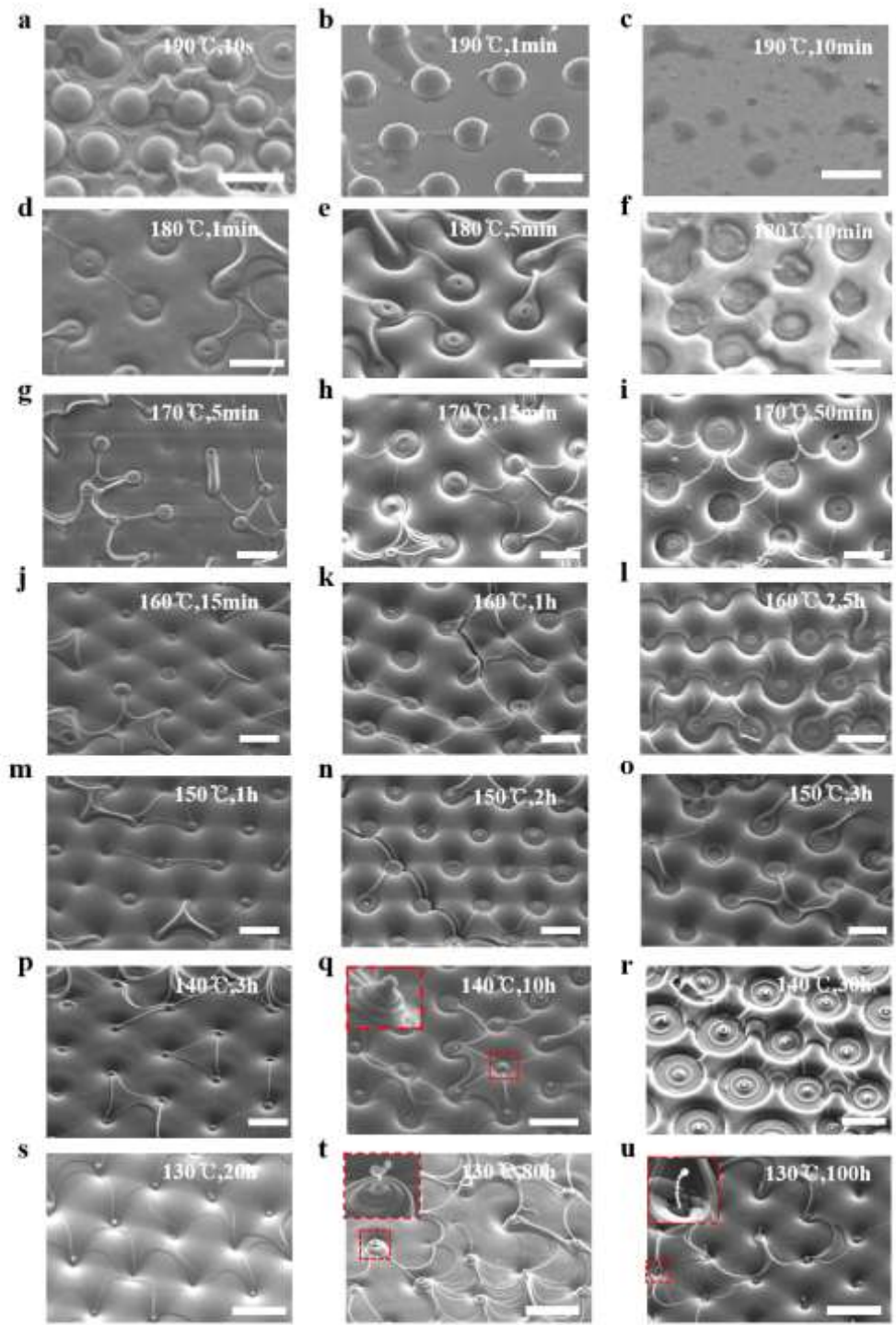

**Figure 4. SEM images corresponding to the sintering diagram** (a-c) Sintering at 190 ℃ for 10s, 1min and 10min, respectively. (d-f) Sintering at 180 ℃ for 1min, 5min and 10min, respectively. (g-i) Sintering at 170 ℃ for 5min, 15min and 50min, respectively. (j-l) Sintering at 160 ℃ for 15min, 1h and 2.5h, respectively. (m-o) Sintering at 150 ℃ for 1h, 2h and 3h, respectively. (p-r) Sintering at 140 ℃ for 3h, 10h and 30h, respectively. (s-u) Sintering at 130 ℃ for 20h, 80h and 100h, respectively. All scale bars are 10 μm.

(130-150) ℃, and low-to-moderate sintering times;

(5) Concentric rings with cusp-like depressions in the centres (Fig.4i,l,o), formed from the conical pyramids at high temperatures as the sintering time increases;

(6) Concentric rings with spherical caps in the centres (Fig. 2b, Fig.4r), formed from the conical pyramids at low temperatures as the sintering time increases;

(7) Udumbara flowers (Fig.2a, Fig.4u), formed at low temperatures after relatively long sintering between $10^5$ and $10^6$ seconds; further sintering results in disappearance of the Udumbara flowers and formation of concentric rings at the substrate, with the Udumbara stalk being reduced to a small cup, Fig.3 and Supplementary Fig.8.

(8) Domes (Fig.4a,b,c), formed at the highest temperatures of 190 ℃.

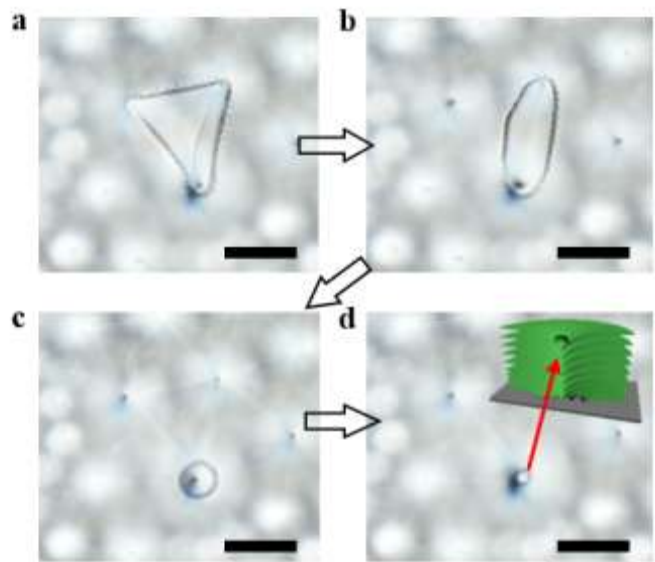

**Figure 5. Spherical cap formation at the TFCD apex during thermal sintering at 130℃** (a) The uppermost layer of the liquid crystal ruptures, forming a shrinking island of a curvilinear triangular shape; (b) the ruptured layer shrinks; (c) the edge of the shrinking layer adopts a circular shape around the central cusp-like depression at the axis of the TFCD; (d) After ~ 30h, the shrinking layer is capped by a spherical surface. The inset in (d) is a schematic sketch of a cross section of the TFCD with a spherical cap. Time interval between each two consecutive images is about 10 h. All scale bars are 10 μm.

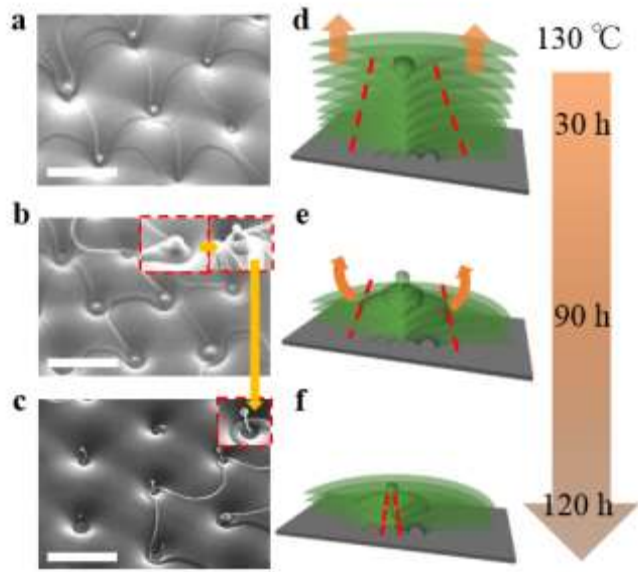

**Figure 6. Formation of the Udumbara flower structure during sintering at 130°C** (a) After 30h sintering, a spherical cap forms at the TFCD centre, following the process illustrated in Fig.5. (b) Further sintering for ~90h removes more material from the periphery of TFCD, leaving conical pyramids at the central parts of TFCDs. (c) Udumbara structures formed after ~120 h of sintering, with the stalks along the straight central defect lines of the original TFCDs. The insets in (b and c) are magnified SEM images, illustrating morphological change from the conical pyramid to the Udumbara structure. All scales are 10 µm. The parts (d,e,f) schematize layers configurations for the corresponding textures in (a,b,c).

**Formation of the Udumbara flower structure** Figure 6 illustrates the development scenario of Udumbara flowers (7), as a function of sintering time. The first stage, Fig.6a,d, is the formation of a spherical cap at the top of the vertical defect line, detailed in Fig.5, through the rupture, shrinkage and closure of the uppermost layer of material. The uppermost layers around the central TFCD region continue to evaporate and shrink with time, leaving remnants of SmA layers around the central defect line; the result is a conical pyramid with a spherical cap at the apex, Fig.6b. The side surface of the pyramid experiences reconstruction: tangential orientation of the molecules at the SmA-air interface associated with the exposed edges of the shrinking layers is being replaced by homeotropic orientation, which necessitates hemi-torical rings, as sketched in Fig.6e. As time elapses, SmA molecules sublimate from both the areas around the pyramid and from the conical side surface of the pyramid, forming a stalk of the Udumbara flower with a spherical cap at the top and hemi-torical rings shrinking into small beads beneath it, Fig.6c,f. If sintering extends beyond 120 h, the Udumbara flower vanishes. Generally, the universal patterns that form immediately prior to complete evaporation of the SmA films after prolonged sintering at moderate temperatures (130-180) ℃, are of the types (5,6), with the concentric rings sharing the centre of the original TFCDs, Fig.3 and Supplementary Fig.8.

## Discussion

According to the Kelvin's equation, the saturated local vapour pressure over a patch of a deformed free surface is determined by its local mean curvature $H$:

$$p = p_0 \left(1 + \frac{2H\sigma_\perp v_1}{k_B T}\right), \quad (1)$$

where $p_0$ is the saturated vapour pressure over the flat surface, $\sigma_\perp$ is the SmA-air surface tension for perpendicular anchoring, $v_1$ is the volume of one molecule, $k_B$ in the Boltzmann constant, $T$ is the absolute temperature. Dependence of $p$ on the mean curvature $H$ is one of the key mechanisms responsible for the observed rich morphology of SmA shape reconstruction during sintering, since the local pressure determines the rate of sublimation/condensation, expressed as the number $n$ of molecules that leave the surface per second per unit area, $n = \frac{\alpha p}{\sqrt{2\pi M k_B T}}$; here $\alpha$ is the evaporation coefficient and $M$ is the molecular weight.

The equilibrium SmA film with a free surface exhibits an array of TFCDs[12,19,29] that satisfy the antagonistic boundary conditions. Packing of equidistant layers within each TFCD forms a cusp-like depression at the apex.[12,14,19,30] We adopt the simplest model of the interface by assuming that it follows the uppermost curved SmA layer; the assumption is justified by Fig.1b, in which there are no edges that might be associated with the layer's ruptures.

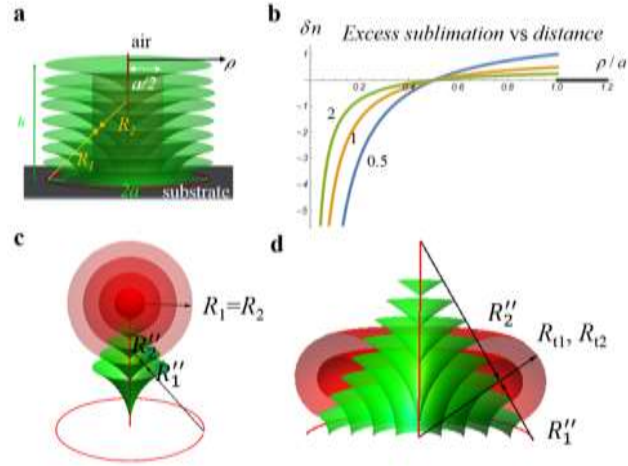

**Figure 7. Morphological reconstructions during thermal sintering of smectic films** (a) The mean curvature within the TFCD changes from negative inside the cylinder of radius $\rho = a/2$ to positive within the region $a/2 < \rho < a$. (b) Excess sublimation rate (arbitrary units) vs. normalized distance $\rho/a$ from the TFCD axis, as predicted by Eq.(2), for three values of $h/a = 0.5; 1; 2$ (shown on the plot); the region $\rho/a > 1$ corresponds to the flat layers of zero curvature located in the interstitials between the TFCDs, with no excess sublimation; the region $0.5 < \rho/a < 1$ corresponds to enhanced sublimation and the region $0 < \rho/a < 0.5$

The SmA-air interface exhibits a variable mean curvature $H = \frac{1}{2}\left(\frac{1}{R_1} + \frac{1}{R_2}\right)$ that depends on the distance $\rho$ (measured along the bottom substrate from the TFCD axis) and the thickness $h$ of the film (measured in the region between the TFCDs), Fig. 7a. The principal radii of the interface on top of a TFCD of a radius $a$, can be expressed in terms of $h$ and $\rho$, as $R_1 = h > 0$ and $R_2 = h - a/\sin\theta$, where $\theta$ is the angle between the normal to the substrate and the director, Fig. 1c. Therefore, the mean curvature $H(\rho,h) = \frac{1}{h}\left(1 - \frac{a}{2\rho}\right)$ changes its sign as a function of $\rho$, being positive at the periphery, $a/2 < \rho \leq a$, and negative closer to the TFCD axis, $0<\rho<a/2.$ In the flat interstitial regions outside the TFCDs, $H=0$. According to the Kelvin's law, variation of the mean curvature produces a variation in the evaporation rates over the SmA-air interface:

$$\delta n = \left(1 - \frac{a}{2\rho}\right) \frac{\sqrt{2}\alpha\sigma_\perp v_1 p_0}{\sqrt{\pi M} h (k_B T)^{3/2}}. \qquad (2)$$

Emission of molecules is enhanced, $\delta n > 0$, at the periphery of each TFCD, where $a/2 < \rho < a$, and suppressed, $\delta n < 0$, near the TFCD centre, at $0 < \rho < a/2$, Fig. 7b. In the flat interstitial regions between the TFCDs, $\delta n = 0$. The regime $\delta n < 0$ represents enhanced capillary condensation, as the lower local pressure drives the molecules emitted elsewhere, towards the TFCD centre. The axis of TFCD is the region of extremely favoured condensation, since for very small distances, $\rho \ll a$, the negative mean curvature becomes very large, $H \to -\dfrac{a}{2\rho h}$.

Equations (1) and (2) are written in the limit of thermodynamic equilibrium, which is not the case of the experiment. Nevertheless, the model helps to understand the counterintuitive experimental features, most importantly, why the thermal treatment leaves the material in the TFCDs' centres and removes it from the TFCD periphery, eventually forming conical pyramids (3,4) and Udumbara flowers (7) centred at the TFCDs axes. Qualitatively, Eq.(2) predicts enhanced sublimation from the periphery of the original TFCDs, $a/2 < \rho < a$, Fig.7a,b, which explains the deep empty pockets around the Udumbara flower stalks, Fig. 2a and conical pyramids, Fig. 2c. Within the central regions, $0 < \rho < a/2$, Fig.7a, Eq.(2) predicts condensation rather than emission, because of the negative mean curvature. In Fig. 7b, the range $0 < \rho/a < 0.5$ corresponds to negative excess sublimation, i.e., condensation. Note the dependence of the function $\delta n(\rho)$ on the ratio $h/a$ in Fig.7b: in shallow films with a small $h/a$, evaporation from the periphery is stronger than in thicker films.

Variation of the local vapour pressure triggers material transfer along the smectic-air interface, supporting re-condensation of the emitted molecules at the central cusp-like depression of a TFCD, $\rho \ll a$, where the negative mean curvature is very large. Condensation assists the formation of spherical caps in the structures of type (2), (4), (6-8), Figs.3-6. The latter is expected to be a complicated dynamic process: Once the spherical cap of positive mean curvature is formed, the local emission rate increases, and the cap could evaporate, leaving a depression behind (structures (1), (3) and 5)). Experimental data suggest that the depressions are favoured at higher temperatures and the spherical caps at lower temperatures; in many cases, these two features coexist, see, for example, Fig.4h, reflecting a fine balance of pressure gradients, emission/condensation rates, diffusion, surface anchoring and bulk elasticity.

Dynamics of spatially variable sublimation, condensation, and restructuring through replacement of unfavourable tangential molecular orientation at the SmA-air interface with a perpendicular orientation implies that the film profile is constantly changing in space and, importantly, time. Consider the scenario in which a pyramid (3,4) is formed, Figs.2c, 6b. As the molecules leave the region $a/2 < \rho < a$, there are two newly formed SmA-air interfaces. One is a cylindrical boundary between the original TFCD and the flat SmA layers outside it, of a radius $\rho = a$. The second newly formed surface is a cylinder of a smaller radius $\rho \approx (1/2)a$, Fig.7a. It encloses the central TFCD region that is a predecessor of the pyramids and Udumbara flowers. Once this cylindrical surface of radius $\rho \approx (1/2)a$ forms, it becomes the site of preferred emission, because of its positive mean curvature that increases from the initial value $H = (1/R_1 + 1/R_2)/2 = 1/(2\rho) \approx 1/a$ to a larger value, as the material evaporates and the remaining central portion of the TFCD shrinks. Sublimation eventually forms a pyramid or an Udumbara stalk, exposing the peripheral parts of TFCD bases and leaving the remnants of SmA in the interstitial regions, where $H = 0$ and $\delta n = 0$. It is expected that some of the emitted molecules can re-condense, helping to produce the circular rings patterns of type (5,6), Fig. 2b. Since the sublimation progresses from the top to the bottom, it is natural to expect that for relatively short sintering times, the intermediate structure would be that of a central conical pyramid, Fig. 2c, 6b (structures of type (3),(4)).

The experiment demonstrates that in all reconstructed morphologies, the orientation of molecules remains perpendicular to the SmA-air interface. This is the case of hemi-toroidal formations of positive Gaussian curvature in Figs. 2b,c, and spheres in Figs. 2a,d. The SmA layers within the TFCDs, spherical and toroidal packings are all of the Dupin cyclides types[2], with focal surfaces in the form of lines (for TFCDs and concentric tori) and points (for spheres). Matching between the remnants of TFCDs and spheres/toroids is facilitated by the geometrical rules according to which the normal vectors to the SmA layers are everywhere straight lines.[2] For example, the normal lines in Fig. 1c join points of the axial defect to the points of the circular TFCD base. These rules are also observed in matching the TFCDs with the spherical packings, Fig. 7c, and toroidal layers, Fig. 7d. In Fig. 7c, the concentric spherical layers have a common centre of curvature at the TFCD apex. The layers from within the TFCD (shown in green) and the spherical (red) part match each other smoothly, as both families of layers are perpendicular to the conical interface separating them, Fig. 7c. This construction, proposed by Sethna and Kleman[31] and confirmed for SmA droplets[28], allows one to match layers with positive and negative Gaussian curvature. Similar consideration applies to toroidal formations, Fig. 7d. In Fig. 7d, the toroidal caps of a positive Gaussian curvature are deposited on the lateral conical surface; other cases, such as toroidal formations around the cylindrical Udumbara flower stalks, Fig. 2a, and relatively flat bases of TFCDs, Fig. 2c, follow the same basic geometry. The regions with a different Gaussian curvature are separated by interfaces that carry an elastic energy.[32] This energy, however, is much smaller than the anisotropic surface anchoring energy gain, as discussed below for the case of hemispherical domes.

The dome caps the circular base of the TFCD, thus replacing the surface energy $\pi\sigma_\parallel a^2$ or $\pi\sigma_\angle a^2$ of the TFCD base exposed to air, with the energy $2\pi\sigma_\perp a^2 + 2\pi(2K+\bar{K})a$. Here $\sigma_\parallel$, $\sigma_\angle$ and $\sigma_\perp$ are the surface tensions for tangential, tilted, and perpendicular orientation of molecules at the SmA-air interface, respectively, $K$ is the splay elastic constant and $\bar{K}$ is the saddle splay elastic constant. The calculation of the elastic energy is based on the standard free energy density of SmA deformations,

$$f = 2KH^2 + \bar{K}G. \qquad (3)$$

For the expected $\sigma_\perp = 2 \times 10^{-2}$ J m$^{-2}$,[33] $a = 5\,\mu m$, $K = 30$ pN (see, for example, [2]), the surface energy of the hemispherical dome, $2\pi\sigma_\perp a^2 \approx 3\times 10^{-12}$ J, is significantly higher than its elastic energy, one the order of $4\pi Ka \approx 10^{-15}$ J; although the saddle-splay modulus $\bar{K}$ is not known for our system, it could not change the balance, as $\bar{K}$ is unlikely to be orders of magnitude larger than $K$.[2] The estimates above show that the anisotropic surface tension is a prevailing factor in shaping the SmA free surface. The surface tension coefficients for unfavourable tangential and tilted orientation should satisfy the condition $\sigma_\parallel, \sigma_\angle > 2\sigma_\perp$ in order for the spherical domes to cap the TFCDs remnants. Although the surface tension of the free SmA surface has been measured only for the case of homeotropic anchoring,[33] it is very likely that the needed condition is satisfied. Tilted or tangential alignment of SmA molecules implies reconstruction of SmA layers through dislocations or through creation of a corrugated profile. For small tilts $\alpha$ from the perpendicular alignment, the surface tension increases as $\sigma \approx \sigma_\perp + K|\alpha|/\lambda$, where $\lambda$ is the length scale on the order of the SmA layer thickness.[34] For the typical $K \sim 30\,\text{pN}$ and $\lambda \sim 3\,\text{nm}$, one estimates $K/\lambda \sim 10^{-2}$ J m$^{-2}$, which is of the same order as $\sigma_\perp = 2 \times 10^{-2}$ J m$^{-2}$ [30]. The conclusion about the prevalence of anisotropic surface anchoring energy gain $\sim (\sigma_\perp - \sigma_\parallel) L^2$ over a patch of area $\sim L^2$, as compared to the typical bulk energy cost, $\sim KL$, remain valid for other geometries of packing, as long as $L > K/(\sigma_\parallel - \sigma_\perp)$, i.e. practically for all scales exceeding the molecular length $\lambda$.

In summary, thermally sintered SmA films produce morphologies with spherical and toroidal elements with positive Gaussian curvature that is rarely met in equilibrium smectic samples. One of the reasons is that $G > 0$ implies an increase of the mean curvature and thus a higher energy cost of packing. The complex morphological changes involve a number of contributing mechanisms, such as sublimation and condensation with the rates that vary in space and time, anisotropic interfacial energies, bulk elasticity and anisotropic diffusion of matter. We outlined how the classic Kelvin equation can contribute to understanding of the key observed features. Much more needs to be done, both experimentally and theoretically, before the morphological dynamics can be described accurately. However, it is already clear that sublimation/condensation of smectic liquid crystals represents a very intriguing field for research and potential applications in terms of submicron-scale manipulation of soft matter.

The presented phenomena introduce a new concept in soft matter design that uses sublimation and condensation as the tool to control curvature of layers and interfaces. The theme is connected to the problem of morphogenesis in living matter. It would be of especial interest to combine the effects of thermal sublimation/condensation with the recently proposed approaches to topological templating[15,17] that can control the density and geometry of focal conic domains.

## Methods

**Synthesis of sublimable liquid crystals**

The LC materials, **Y001 – Y006** were prepared as previously reported.[19] Phase transitions for the samples were characterized by differential scanning calorimetry (Q1000 V9.9 Build 303).

**Fabrication of morphologies by thermal sintering**

Silicon wafers were cleaned using acetone and methanol, followed by rinsing with deionized water. The cleaned silicon substrates were spin-coated with polyethyleneimine (PEI; Aldrich; MW: 60,000) for tangential anchoring of the LC. A crystalline powder of **Y002** at the silicon substrate was heated to the isotropic phase (200°C) on a hot stage (LINKAM LTS350) regulated by a temperature controller (LINKAM TMS94). The sample was cooled to 190°C at a rate of 10 °C min$^{-1}$. The sublimation experiments were performed in the range between 190°C and 130°C with an increment of 10 °C for different sintering times, ranging from 5 to $10^6$ seconds. Thermal treatment was performed in the air. The instantaneous rates for mass loss percentage at the temperatures 190 °C, 180 °C, 160 °C and 130 °C, were determined to be 1.64, 0.82, 0.30 and 0.02 %/℃, respectively, Supplementary Fig. 9. The volume available for the SmA molecules to sublimate is orders of magnitudes larger than the volume of the original

SmA film (cubic decimetres vs cubic millimetres). Following the thermal treatment, the samples were cooled down to the room temperature, preserving the SmA structure (as verified by X-ray diffraction). The surface morphology was visualized by sputter-coating platinum (Pt) film and imaging the latter under a scanning electron microscope (SEM).

**Imaging morphologies of thermally treated samples**

The thermally annealed samples were coated with a 5 nm thick layer of Pt and then observed by a field emission SEM (FE-SEM; Hitachi, S-4800) with 7kV and 7μA in high vacuum. Surface topography measurements were performed using an AFM (Bruker, Multimode-8) equipped with a 100 μm$^2$ scanner in tapping mode under ambient conditions and an antimony doped silicon cantilever with a spring constant of 20~80N/m and frequency($f_0$) of 300 kHz was used.

# References

1    Bourgine, P. & Lense, A. *Morphogenesis : origins of patterns and shapes*. Vol. xvii (New York: Springer Verlag, 2011).

2    Kleman, M., & Lavrentovich, O. D. *Soft Matter Physics: An Introduction*. (New York : Springer, 2003).

3    Friedel, G. & Grandjean, F. Observations géométriques sur les liquides á conique focales. *Bull Soc Fr Minéral* **33**, 409-465, (1910).

4    Lavrentovich, O. D. Filling of Space by Flexible Smectic Layers. *Mol. Cryst. Liq. Cryst.* **151**, 417-424, (1987).

5    Adamczyk, A. Phase-Transitions in Freely Suspended Smectic Droplets - Cotton-Mouton Technique, Architecture of Droplets and Formation of Nematoids. *Mol. Cryst.*

*Liq. Cryst.* **170**, 53-69, (1989).

6    Arora, S. L., Palffy-Muhoray, P., Vora, R. A., David, D. J. & Dasgupta, A. M. Reentrant Phenomena in Cyano Substituted Biphenyl Esters Containing Flexible Spacers. *Liq. Cryst.* **5**, 133-140, (1989).

7    Pratibha, R. & Madhusudana, N. V. Cylindrical Growth of Smectic-A Liquid-crystals from the Isotropic-Phase in some Binary-Mixtures. *J. Phys. II* **2**, 383-400, (1992).

8    Fournier, J. B. & Durand, G. Focal Conic Faceting in Smectic-A Liquid Crystals. *J. Phys. II* **1**, 845-870, (1991).

9    Naito, H., Okuda, M. & Ou-Yang, Z.-c. Preferred equilibrium structures of a smectic-Aphase grown from an isotropic phase: Origin of focal conic domains. *Phys. Rev. E* **52**, 2095-2098, (1995).

10   Blanc, C. & Kleman, M. The confinement of smectics with a strong anchoring. *Eur. Phys. J. E* **4**, 241-251, (2001).

11   Iwashita, Y. & Tanaka, H. Spontaneous Onion-Structure Formation from Planar Lamellar Nuclei. *Phys. Rev. Lett.* **98**, 145703, (2007).

12   Fournier, J. B., Dozov, I. & Durand, G. Surface frustration and texture instability in smectic-Aliquid crystals. *Phys. Rev. A* **41**, 2252-2255, (1990).

13   Guo, W. & Bahr, C. Influence of anchoring strength on focal conic domains in smectic films. *Phys. Rev. E* **79** 011707, (2009).

14   Kim, Y. H. *et al.* Confined Self-Assembly of Toric Focal Conic Domains (The Effects of Confined Geometry on the Feature Size of Toric Focal Conic Domains). *Langmuir* **25**, 1685-1691, (2009).

15   Honglawan, A. *et al.* Pillar-assisted epitaxial assembly of toric focal conic domains of smectic-a liquid crystals. *Adv. Mater.* **23**, 5519-5523, (2011).

16    Honglawan, A. *et al.* Topographically induced hierarchical assembly and geometrical transformation of focal conic domain arrays in smectic liquid crystals. *Proc. Natl. Acad. Sci. USA* **110**, 34-39, (2013).

17    Beller, D. A. *et al.* Focal Conic Flower Textures at Curved Interfaces. *Phys. Rev. X* **3**, 041026, (2013).

18    Ohzono, T., Takenaka, Y. & Fukuda, J.-i. Focal conics in a smectic-A liquid crystal in microwrinkle grooves. *Soft Matter* **8**, 6438-6444, (2012).

19    Yoon, D. K. *et al.* Internal structure visualization and lithographic use of periodic toroidal holes in liquid crystals. *Nat. Mater.* **6**, 866-870, (2007).

20    Zappone, B., Meyer, C., Bruno, L. & Lacaze, E. Periodic lattices of frustrated focal conic defect domains in smectic liquid crystal films. *Soft Matter* **8**, 4318-4326, (2012).

21    Coursault, D. *et al.* Linear self-assembly of nanoparticles within liquid crystal defect arrays. *Adv. Mater.* **24**, 1461-1465, (2012).

22    Zappone, B. *et al.* Self-ordered arrays of linear defects and virtual singularities in thin smectic-A films. *Soft Matter* **7**, 1161-1167, (2011).

23    Kim, Y. H., Yoon, D. K., Jeong, H. S., Lavrentovich, O. D. & Jung, H.-T. Smectic Liquid Crystal Defects for Self-Assembling of Building Blocks and Their Lithographic Applications. *Adv. Func. Mater.* **21**, 610-627, (2011).

24    Yoon, D. K. *et al.* Three-dimensional textures and defects of soft material layering revealed by thermal sublimation. *Proc. Natl. Acad. Sci. USA* **110**, 19263-19267, (2013).

25    Wakai, F. & Aldinger, F. Sintering through surface motion by the difference in mean curvature. *Acta Materialia* **51**, 4013-4024, (2003).

26    Herring, C. Effect of Change of Scale on Sintering Phenomena. *J. Appl. Phys.* **21**, 301-303, (1950).

27  Dalal, S. S., Walters, D. M., Lyubimov, I., de Pablo, J. J. & Ediger, M. D. Tunable molecular orientation and elevated thermal stability of vapor-deposited organic semiconductors. *Proc. Natl. Acad. Sci. USA* **112**, 4227-4232, (2015).

28  Lavrentovich, O. D. Hierarchy of Defects on Filling up of Space by Flexible Smectic-A Layers. *Zhurnal Eksperimentalnoi Teor. Fiz.* **91**, 1666-1676/English translation: *Sov. Phys. JETP* **64**, 984-990, (1986).

29  Harth, K., Schulz, B., Bahr, C. & Stannarius, R. Atomic force microscopy of menisci of free-standing smectic films. *Soft Matter* **7**, 7103-7111, (2011).

30  Guo, W. & Bahr, C. Influence of phase sequence on focal conic domains in smectic films. *Phys. Rev. E* **79**, 061701, (2009).

31  Sethna, J. P. & Kléman, M. Spheric domains in smectic liquid crystals. *Phys. Rev. A* **26**, 3037-3040, (1982).

32  Kleman, M. & Lavrentovich, O. D. Grain boundaries and the law of corresponding cones in smectics. *Eur. Phys. J. E* **2**, 47-57, (2000).

33  Lavrentovich, O. D. & Tarakhan, L. N. Temperature dependence of the surface tension at the liquid crystal-isotropic fluid interface. *Poverkhnost (=Surface in Russian)* **1**, 39-44, (1990).

34  Durand, G. Recent advances in nematic and smectic A anchoring on amorphous solid surfaces. *Liquid Crystals* **14**, 159-168, (1993).

# Supplementary Figures 1-9 and Table 1

**Supplementary Fig. 1**

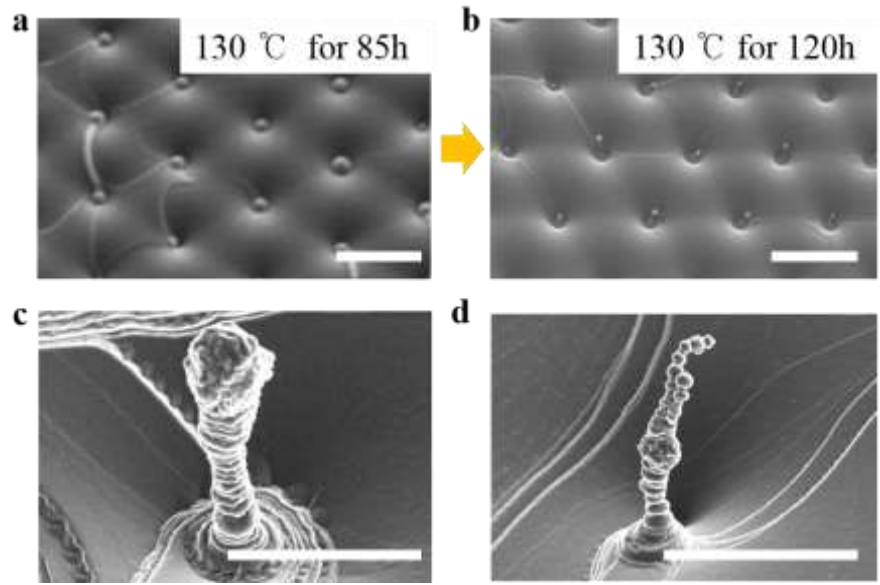

**Supplementary Fig. 1**. (a,b) SEM images of Udumbara structures formed by thermal annealing at low temperature (130 ℃). (c,d) SEM images of Udumbara morphologies of Y003 emerging during thermal annealing at 130 ℃. All scale bars are 10μm.

**Supplementary Fig. 2.**

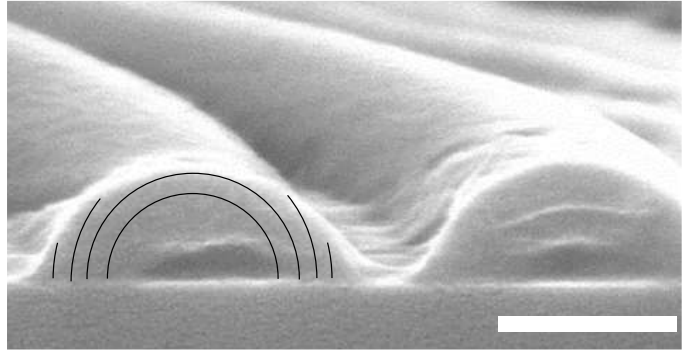

**Supplementary Fig. 2.** The cross sectional SEM image of rings within the TFCD base formed by sintering at 160 ℃ for 40 min. The sample was cut by glass cutter at room temperature. The shape is not exactly semitoroidal, which can be related either to modified surface anchoring or to the presence of incomplete smectic layers, as shown schematically on the left hand side. The scale bar is 200 nm.

**Supplementary Fig. 3**

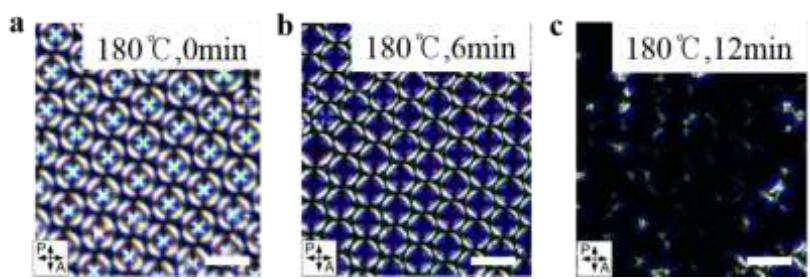

**Supplementary Fig. 3**. Polarizing optical textures of TFCD film subject to thermal sublimation at 180℃ as a function of time. (a) 0 min, (b) 6min, (c) 12min. All scale bars are 10μm.

**Supplementary Fig. 4**

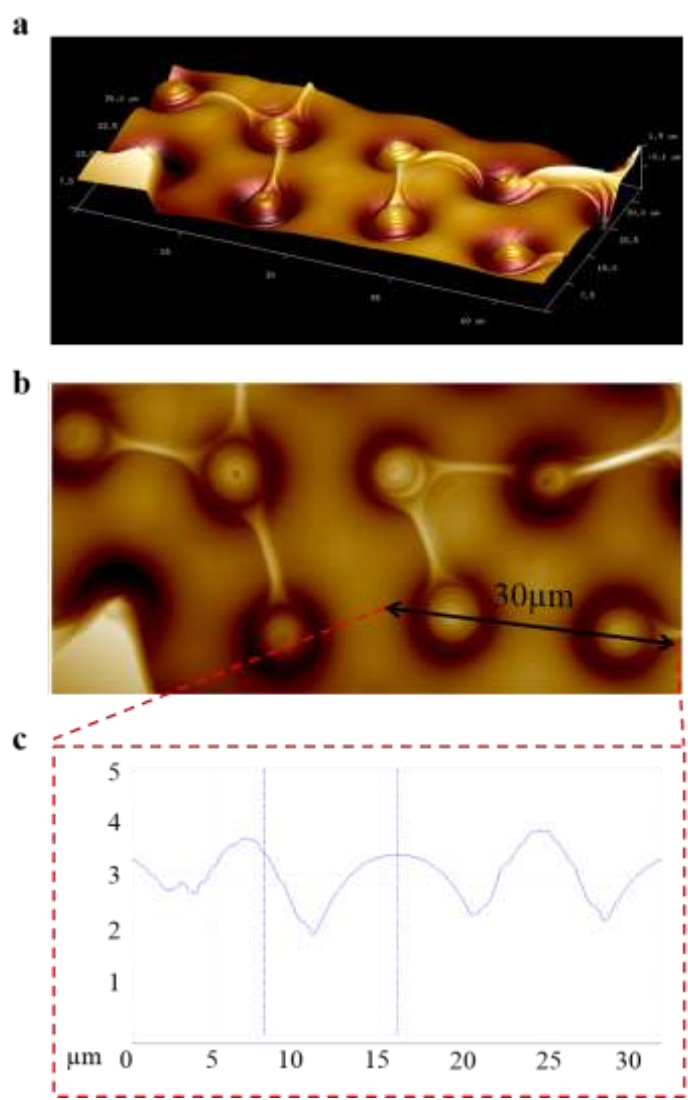

**Supplementary Fig. 4**. Topographical structure analysis of conical pyramid-like textures formed from TFCD arrays by sintering at 180 °C for 10 min. The sample was analysed at room temperature.

**Supplementary Fig. 5**

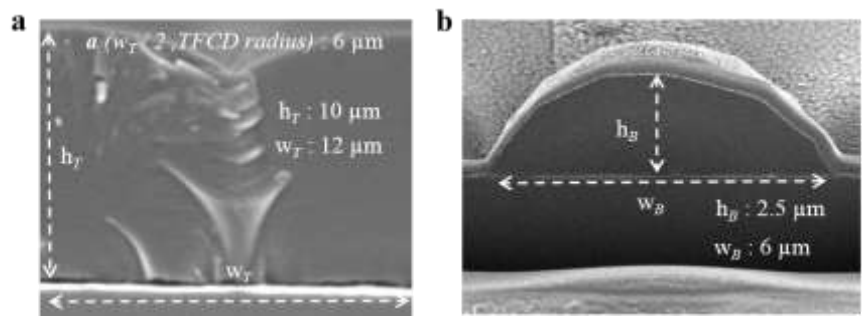

**Supplementary Fig. 5.** The cross sectional SEM images of the original TFCD and dome structure that occurs after thermal treatment. (a) The volume of TFCD ($V_T$) is about 1130 μm³ and the volume of dome structure ($V_B$) ~ 113 μm³, showing the evaporated volume ratio ($V_T - V_B$) / $V_T$ = 0.9.

**Supplementary Fig. 6**

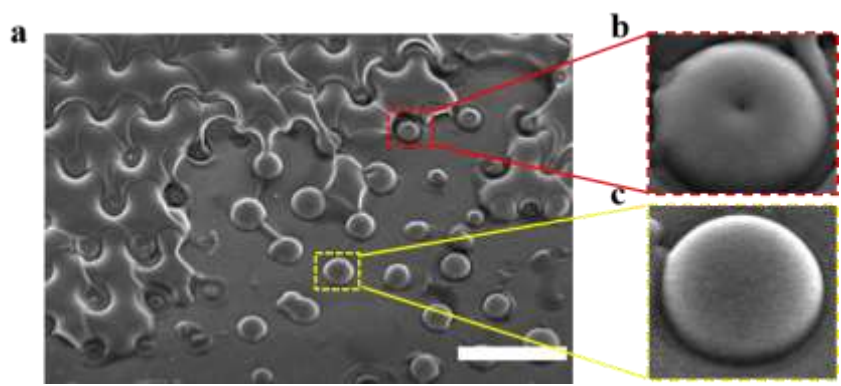

**Supplementary Fig. 6. (a)** SEM images of SmA film sintered at 190 ºC (the scale bar is 10 μm) with **(b)** toroidal and **(c)** spherical shape replacing the original TFCDs.

**Supplementary Fig. 7**

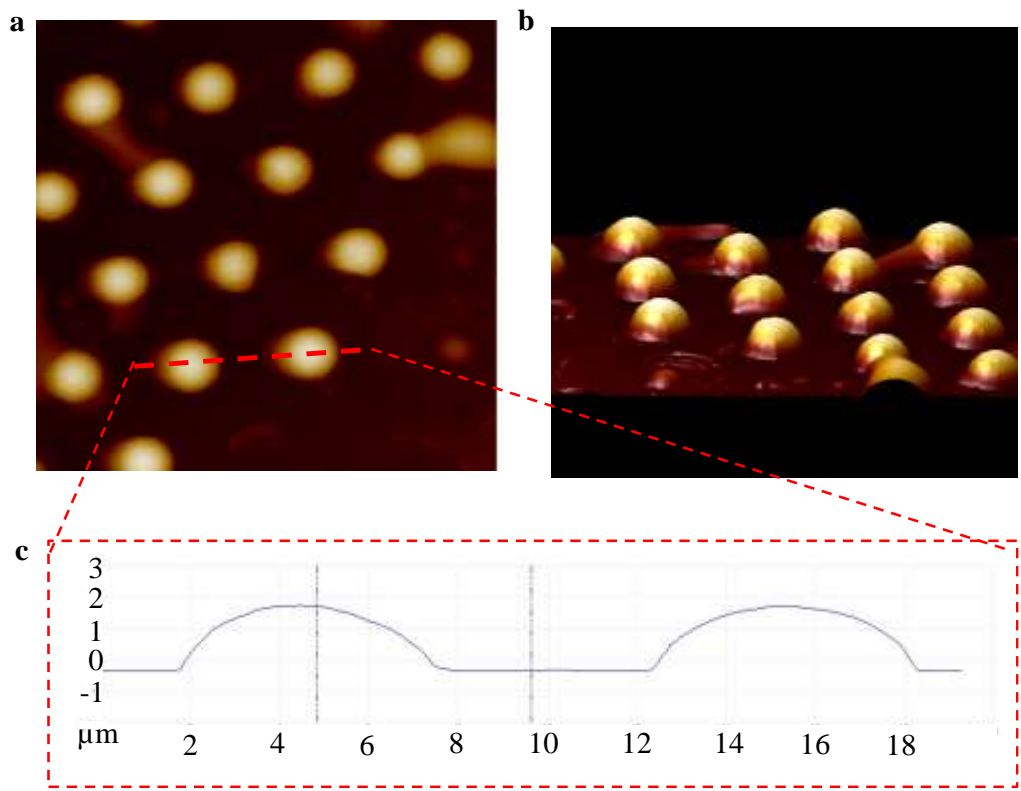

**Supplementary Fig. 7**. AFM textures of dome structures formed from TFCD by sintering at 190 °C for 5 min. The measurements are performed at room temperature.

**Supplementary Fig. 8**

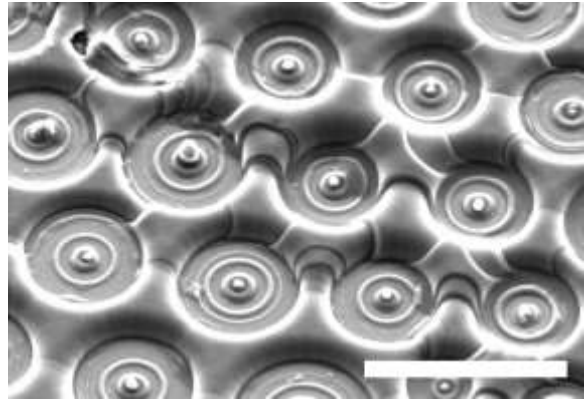

**Supplementary Fig. 8.** SEM texture of a concentric rings structure formed from an Udumbara structure after a prolonged 200h sintering of the smectic film at 130 °C. The Udumbara stalk leaves a small piece of the material at the substrate. The scale bar is 20 µm.

**Supplementary Fig. 9**

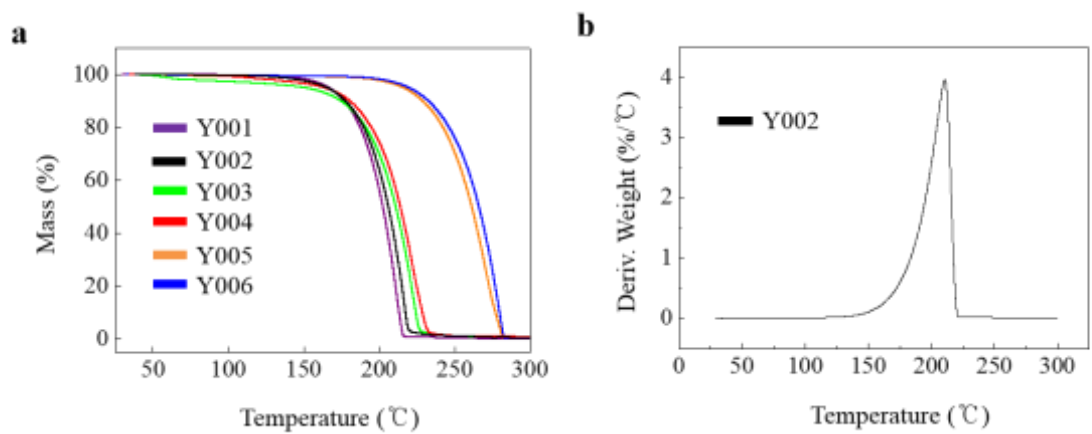

**Supplementary Fig. 9.** (a) Thermogravimetry analysis (TGA) plots for Y001-Y006 as a function of temperature during heating (heating rate: 1°C/min) shows that sublimation of Y001-Y004 occurred before the material reached their respective isotropic temperatures; Y005, Y006 show very weak sublimation in the SmA phase. (b) The derivative weight percentage of Y002 as a function of temperature presents the sublimation rate.

**Supplementary Table 1.**

| | Cooling | | | | | | |
|---|---|---|---|---|---|---|---|
| **Y001** | I | 207 | SmA | 146 | SmE | 69 | Cr |
| **Y002** | I | 188 | SmA | 111 | SmE | 57 | Cr |
| **Y003** | I | 163 | SmA | 63 | SmE | 48 | Cr |
| **Y004** | I | 154 | SmA | 59 | SmE | 42 | Cr |
| **Y005** | I | 158 | SmA | 68 | SmE | 51 | Cr |
| **Y006** | I | 149 | SmA | 69 | SmE | 56 | Cr |

**Supplementary Table 1.** Phase transition temperatures of the materials during cooling with the rate 5 ℃ min$^{-1}$ (Cr=crystal, SmE=smetic E, SmA=smectic A, I=isotropic phase).